\documentclass[12pt]{article}
\usepackage[centertags]{amsmath}
\usepackage{amsfonts}
\usepackage{amssymb}
\usepackage{color}
\usepackage{amsthm} 
\usepackage{newlfont}
\usepackage{epsfig}
\usepackage{amscd}
\usepackage{graphicx}
\usepackage{epsfig}
\usepackage{amscd}

\begin{document}

\begin{center}

{\Large \bf The multiple quantum NMR dynamics in systems of equivalent spins with the dipolar ordered initial state}



\vspace{0.3cm}

S.\,I.\,Doronin, E.\,B.\,Fel'dman and A.\,I.\,Zenchuk



Institute of Problems of Chemical Physics, Russian Academy of Sciences,
Chernogolovka, Moscow reg., 142432, Russia
\end{center}


\begin{abstract}
The multiple quantum (MQ) NMR dynamics in the system of equivalent spins with the dipolar ordered initial state is considered. The high symmetry of { the  Hamiltonian responsible for the MQ NMR dynamics (the MQ Hamiltonian)} is used in order to develop the analytical and numerical methods for an investigation of the MQ NMR dynamics in the systems consisting of hundreds of spins from "the first principles". We obtain 
the dependence of the intensities of the  MQ NMR coherences on their orders (profiles of the MQ NMR coherences) for the systems of $200 - 600$ spins. It is  shown that these profiles may be well  approximated  by the exponential distribution functions. We also compare the MQ NMR dynamics in the systems of equivalent spins having two different initial states, namely   the dipolar ordered state and  the thermal equilibrium state in the strong external magnetic field.
\end{abstract}




\section{Introduction}
The multiple quantum (MQ) NMR dynamics in solids \cite{BMGP} is extremely usefull  for an investigation of   solid structures  and dynamical processes therein, for a counting  the number of spins in impurity clusters \cite{BGPGR,H}, for a simplification of the standard NMR spectra \cite{WWP}. Usually the MQ NMR experiments deal with  samples  where the nuclear spin system is initially prepared  in the thermal equilibrium state in the strong external magnetic field  \cite{BMGP}. However, it is possible to carry out the MQ NMR experiments with samples prepared in different initial states \cite{FG}. In particular, one can prepare a spin system in the dipolar ordered state \cite{G} using either the adiabatic demagnetization
method in the rotating reference frame (RRF) \cite{G,SH} or the two-pulse Broekaert-Jeener
sequence \cite{G,JB}. The MQ NMR dynamics with this initial state in small spin systems  has been simulated in refs. \cite{DFKFG,DFKFG2}. Using the dipolar ordered initial state  in the MQ NMR experiment  one should  expect  the earlier  appearance of multiple spin clusters  and correlations in comparison with the MQ NMR experiment with the thermal equilibrium initial state in the strong external magnetic field. In fact, the analysis of the MQ NMR experiments in the six-eight spin systems  \cite{DFKFG,DFKFG2} demonstrates that the six-eight order  MQ coherences appear earlier  in 
the  experiment with the dipolar ordered initial state.

One of the basic problems  for the theoretical description of the MQ NMR experiments is an exponential growth of the density matrix dimensionality  with the increase in the number of spins.
Therefore the modern numerical methods have been developed for simulation of the MQ NMR dynamics, which are based  either on Chebyshev polynomial expansion \cite{DRKN} or quantum parallelism \cite{ZCAPCDRV}. However, these methods  allow one to study  the MQ NMR dynamics in systems of no more than several tens of spins.   A significant progress in this direction has been achieved  in simulation of  the MQ NMR dynamics  in the system of equivalent spins \cite{BKHWW,FR,DFFZ}, which  may be prepared, for instance, filling the closed nanopore with the gas of spin-carring 
molecules (or atoms). 
 The matter is that, due to the special symmetry  of the Hamiltonian governing the dynamics in such spin system, it becomes possible to study the MQ NMR dynamics in systems of hundreds of spins and even more \cite{BKHWW,FR}.  The nature of the above mentioned symmetry  can be clarified as follows.  { As far as the characteristic time between two successive collisions with the nanopore walls is} several orders less that the time of mutual flip-flops of  any two nuclear spins (which is defined by their dipole-dipole interaction (DDI)) \cite{BKHWW,FR}, it seems to be reasonable to use the averaged DDI, which  may be obtained by averaging over the spin positions in the nanopore \cite{BKHWW}. This means that the constant of the averaged DDI remains the same for any pair of spins in the nanopore {\cite{BKHWW}},
so that the nuclear spins become equivalent.  For this reason, the Hamiltonian of the nuclear spin DDI in the nanopore commutes with the operator of the square of the
 total spin angular momentum $I^2$ \cite{DFFZ,DFFZ2}. Thus, it becomes possible to use the basis of the common eigenfunctions for the operator of the square of the
 total spin momentum $I^2$ and of its projection $I_z$ on the direction of the external magnetic field instead of the standard multiplicative basis of the eigenfunctions of $I_z$, which yields  an exponential growth of the Hilbert space dimensionality  with the increase in the  number of spins \cite{DFFZ,DFFZ2}. As a  result, { we simplify calculations which allows us to succeed in}  both  the investigation of the MQ NMR dynamics  in systems of 200 -- 600 spin-1/2 particles \cite{DFFZ} and  in the study of the dependence of the coherence  relaxation time on the MQ NMR  coherence order and the  number of spins \cite{DFZ}.  Emphasize that  the nuclear spin system with the thermodynamic equilibrium  initial state in the strong external magnetic field  is used in refs.\cite{DFFZ,DFFZ2,DFZ}. 

The MQ NMR dynamics in the large system of equivalent spins with the dipolar ordered initial state  is studied in the present paper. 
 The theory of MQ NMR dynamics of equivalent spins with this  initial state  is given in Sec.\ref{Section:theory}.  The dependence of the MQ NMR coherence intensities on the coherence orders (the profiles of MQ NMR coherence intensities) for systems  of 200-600 spins  is represented in  Sec.\ref{Section:numerics}.  The MQ NMR dynamics in systems with the
dipolar ordered initial state is compared with the dynamics in systems with the thermal equilibrium initial state in the strong external magnetic field  in Sec.\ref{Section:comp}.  The basic results  are collected in Sec.\ref{Section:conclusion}.

\section{The MQ NMR coherence intensities in systems of equivalent spins prepared in the dipolar ordered initial state}
\label{Section:theory}

We consider the system of equivalent spin-1/2 particles  with the dipole-dipole interaction (DDI) in the strong external magnetic field. The secular part of the DDI Hamiltonian \cite{G} reads:
\begin{eqnarray}
\label{Hdz}
H_{dz} = \sum_{j<k} D_{jk}(2 I_{jz} I_{kz} -  I_{jx} I_{kx} -I_{jy} I_{ky}),
\end{eqnarray} 
where $D_{jk}=\frac{\gamma^2 \hbar}{2 r_{jk}^3} ( 1-3 \cos^2\theta_{jk})$ is the 
constant of DDI, $\gamma$ is the gyromagnetic ratio, $r_{jk}$ is the distance between 
the $j$th and $k$th spins, $\theta_{jk}$ is the angle between the internuclear vector 
$\vec r_{jk}$ and the external magnetic field $\vec B_0$, and $I_{j\alpha}$ 
($\alpha=x,y,z$)
is the $j$th spin projection  operator on the axis $\alpha$. 
Using either  the adiabatic demagnetization in the rotating reference frame \cite{G,SH} or the Broekaert-Jeener two-pulse sequence \cite{G,JB} one can prepare the spin system in the dipolar ordered initial state  with the following density matrix:
\begin{eqnarray}\label{HT}
\rho(0) = \frac{1}{Z} \exp(-\beta H_{dz}) \approx \frac{1}{2^N} ( 1 -\beta H_{dz}),
\end{eqnarray}
where $\beta=\frac{\hbar}{kT}$ is the inverse  spin temperature ($k$ is the Boltsman constant, $T$ is the temperature), $Z={\mbox{Tr}} \exp(-\beta H_{dz})$ is the partition function and $N$ is the  number of spins. 

If the system under consideration consists of the spin carring molecules (atoms) in the closed nanopore, then    DDI is averaged  (incompletely) by the fast molecular diffusion so that the constants  of DDI for any spin pair equal each other, i.e. $D_{jk}\equiv D$  \cite{BKHWW,FR}. As a  result,  the Hamiltonian (\ref{Hdz}) may be written as follows \cite{FR}:
\begin{eqnarray}\label{Hdz:equiv}
\bar H_{dz} =\frac{D}{2} (3 I_z^2 - I^2),
\end{eqnarray}
where $I_\alpha=\sum_{i=1}^N I_{i\alpha}$, ($\alpha=x,y,z$), $I^2=I_x^2+I_y^2+I_z^2$ is the square of the spin angular momentum.
It is important to justify that the high temperature approximation (\ref{HT})  is applicable to the system of equivalent spins.
The simple analisis \cite{D} demonstrates that approximation (\ref{HT}) for the system with the Hamiltonian (\ref{Hdz:equiv}) is valid  if
\begin{eqnarray}\label{HTc}
\beta D N \ll 1.
\end{eqnarray}
The systems considered in our work have $N=200-600$, so that the condition (\ref{HTc}) is satisfied.

\begin{figure*}
   \epsfig{file=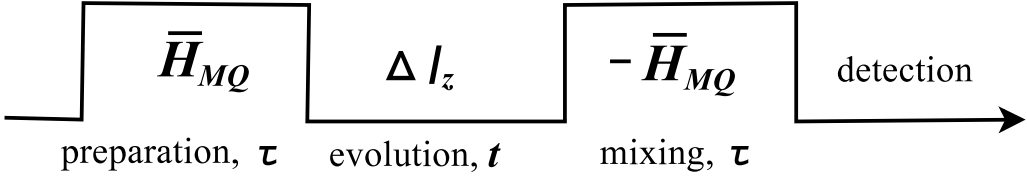
   , scale=0.5,angle=0
}
  \caption{The basic scheme of the four period MQ NMR experiment.
 }
  \label{Fig:NMR} 
\end{figure*}

The MQ NMR experiment consists of four basic periods shown in Fig.\ref{Fig:NMR}: the preparation period  $\tau$, the evolution period $t$, the mixing period $\tau$ and the detection. The MQ NMR coherences are generated on the preparation period due to the irradiation of the sample by the multiple eight-pulse sequence of the resonance pulses \cite{BMGP}. Let us express MQ NMR coherences in terms of the density matrix of the preparation period. 
For this purpose note that the averaged Hamiltonian $\bar H_{MQ}$ (nonsecular two-spin/two-quantum Hamiltonian \cite{BMGP}) describing the MQ dynamics in the system of equivalent spins  during the preparation period in the rotating reference frame   may be written as follows \cite{DFFZ,DFFZ2}:
\begin{eqnarray}\label{HMQ}
\bar H_{MQ} =- \frac{D}{4} \{ (I^+)^2 + (I^-)^2\},
\end{eqnarray}
where $I^+$ and $I^-$ are the raising and lowering operators ($I^\pm=I_x \pm i I_y$).
In order to investigate the MQ NMR dynamics, one has to find the density matrix $\rho(\tau)$ for the spin system solving the Liouville equation \cite{G}:
\begin{eqnarray}
\label{Liouville}
i \frac{d\rho}{d\tau} = [\bar H_{MQ},\rho(\tau)],
\end{eqnarray}
with the initial condition $\rho(0)=\bar H_{dz}$. This  initial condition is obtained from eqs. (\ref{HT}) and (\ref{Hdz:equiv}) by dropping both the unit operator and the factor ($-\beta/2^N$), which are not significant for the MQ NMR dynamics. 
Taking into account  the Hamiltonians on the different periods of the MQ NMR experiment (these Hamiltonians are shown in Fig.\ref{Fig:NMR}) one can write the expression for the dipolar energy $\langle \bar H_{dz} \rangle(\tau,t)$ after the mixing period of the  MQ NMR experiment  (Fig.\ref{Fig:NMR})
as follows:
\begin{eqnarray}\label{H_dz}
\langle \bar H_{dz}\rangle(\tau,t)&=&\frac{{\mbox{Tr}}\{U^+(\tau) e^{-i\Delta t I_z}  U(\tau) 
\bar H_{dz} U^+(\tau) 
e^{i\Delta t I_z} U(\tau) \bar H_{dz}\}}{{\mbox{Tr}}\{ \bar H_{dz}^2\}}=\\\nonumber
&&
\frac{{\mbox{Tr}}\{ e^{-i\Delta t I_z}  \rho(\tau) 
e^{i\Delta t I_z} \rho(\tau)\}}{{\mbox{Tr}}\{ \bar H_{dz}^2\}},
\end{eqnarray}
where { $\rho(\tau)=U(\tau) \bar H_{dz} U^+(\tau)$} is the solution to Eq.(\ref{Liouville}) and $U(\tau)=\exp(-i\bar H_{MQ} \tau)$. 
It is convenient to represent the solution to Eq.(\ref{Liouville}) as the  series \cite{FL}
\begin{eqnarray}\label{rhok}
\rho(\tau) =\sum_k\rho_k(\tau),
\end{eqnarray}
where { $\rho_k(\tau)$ obeys the relationship $[I_z,\rho_k(\tau)]=k\rho_k(\tau)$ which can be considered as the definition of $\rho_k(\tau)$ (in other words, $\rho_k(\tau)$  collects those entries of the density matrix $\rho(\tau)$ which are responsible for the $k$th order MQ NMR coherence)}. Then eq.(\ref{H_dz}) reads
\begin{eqnarray}\label{H_dz2}
\langle \bar H_{dz}\rangle(\tau)=\sum_k e^{-ik\Delta t}\frac{{\mbox{Tr}}\{
\rho_k(\tau)\rho_{-k}(\tau)
\}}{
{\mbox{Tr}} \{ \bar H_{dz}^2\}}= \sum_k e^{-ik\Delta t} J_k(\tau),
\end{eqnarray}
where  the $k$th order MQ NMR coherence intensity $J_k$ is defined as follows \cite{FL}:
\begin{eqnarray}\label{Jk}
J_k(\tau)=\frac{
{\mbox{Tr}} \{ \rho_k(\tau)\rho_{-k}(\tau)\}}{
{\mbox{Tr}} \{ \bar H_{dz}^2\}
}.
\end{eqnarray} 
This formula will be used in the numerical analysis of the MQ NMR coherences  in Sec.\ref{Section:numerics}.

It is easy to write the explicit expression for ${\mbox{Tr}} \{ \bar H_{dz}^2\}$.
First of all, using eq. (\ref{Hdz:equiv}) we may write
\begin{eqnarray}\label{tr}
&&
{\mbox{Tr}} \{ \bar H_{dz}^2\} =\frac{D^2}{4} {\mbox{Tr}}(3 I_z^2 -I^2)^2 =\\\nonumber
&&
\frac{D^2}{4} {\mbox{Tr}}\{
9 I_z^4 - 3 I_z^2 I^2 + I^2 ( I^2 -3 I_z^2)
\}=\\\nonumber
&&
\frac{3D^2}{4} {\mbox{Tr}}\{
3 I_z^4 -  I_z^2 I^2 
\}=\frac{3D^2}{2} {\mbox{Tr}}\{
 I_z^4 -  I_x^2 I_z^2 
\}.
\end{eqnarray} 
Here we  take into account that 
\begin{eqnarray}\label{tr2}
{\mbox{Tr}}\{I^2 ( I^2 -3 I_z^2)\}=0, \;\; {\mbox{Tr}}\{I_x^2 I_z^2\} = {\mbox{Tr}}\{I_y^2 I_z^2\}.
\end{eqnarray} 
Next, it is simple to obtain the explicit expressions for ${\mbox{Tr}} \{I_z^4\}$ and 
${\mbox{Tr}}\{ I_x^2 I_z^2 \}$:
\begin{eqnarray}\label{tr3}
{\mbox{Tr}}\{ I_z^4\} = 2^{N-4} N (3 N -2),\;\;\;
 {\mbox{Tr}} \{I_x^2 I_z^2\}  = 2^{N-4} N^2.
\end{eqnarray}
Finally, eqs.(\ref{tr}-\ref{tr3}) yield the following result:
\begin{eqnarray}
{\mbox{Tr}} \{ \bar H_{dz}^2\} =3 N(N-1) 2^{N-4} D^2.
\end{eqnarray}
Taking into account the structure of the MQ Hamiltonian $\bar H_{MQ}$, it may be readily shown that only the even order MQ NMR coherences appear in our numerical experiment and the coherence order may not exceed the  number of spins $N$ \cite{BMGP,FL}.

It is obvious that the MQ Hamiltonian (\ref{HMQ})  commutes with the square of the  total spin momentum operator  $I^2$ \cite{DFFZ,DFFZ2,DFZ}. As far as $[I^2,I_z]=0$, it is possible to use the basis of the common eigenfunctions of operators $I^2$ and $I_z$ for the description of the MQ NMR dynamics. 
Namely this fact allows one to { avoid} the problem of the exponential growth of the matrix dimensionality with the increase in the  number
of spins, which appears in  the 
traditional multiplicative basis  \cite{BMGP} of the eigenfunctions of the operator $I_z$.  In the new basis, the MQ Hamiltonian $\bar H_{MQ}$ consists of blocks $\bar H_{MQ}^S$ corresponding to the different values of the total spin momentum $S$ ($\hat S^2 = S(S+1)$, $S=\frac{N}{2}, \frac{N}{2}-1, \frac{N}{2}-2,\cdots, \frac{N}{2}-\left[\frac{N}{2}\right]$, $[i]$ is an integer part of $i$):
\begin{eqnarray}
\bar H_{MQ}={\mbox{diag}}\{ \bar H^{\frac{N}{2}}_{MQ} ,\bar H^{\frac{N}{2}-1}_{MQ},
\dots ,\bar H^{\frac{N}{2}-\left[\frac{N}{2}\right]}_{MQ}\}.
\end{eqnarray}
As far as the Hamiltonian $\bar H_{dz}$ (\ref{Hdz:equiv}) is diagonal in the basis of the common eigenfunctions of the operators $I^2$ and $I_z$, then the density matrix $\rho(0)$ may be also splitted into the diagonal blocks $\rho^S(0)$ with $S=\frac{N}{2}, \frac{N}{2}-1, \frac{N}{2}-2,\cdots, \frac{N}{2}-\left[\frac{N}{2}\right]$. Consequently, the matrix $\rho(\tau)$  has also the diagonal block structure with blocks $\rho^S(\tau)$. Thus, the problem becomes separated  into the set of independent problems for each $(2S+1)\times (2S+1)$-dimensional block $\rho^S(\tau)$ which is a solution to the Liouville equation  (\ref{Liouville})  with the Hamiltonian $\bar H_{MQ}^S$. Of course, expansion (\ref{rhok})  may be applied to each block $\rho^S$. The contribution $J_{k,S}$   from the block $\rho^S$ to the intensity  $J_k$  of the $k$th 
order coherence  is defined by the obvious formula \cite{DFFZ}:
\begin{eqnarray}\label{JkS}
J_{k,S}(\tau)=\frac{
{\mbox{Tr}}\{
\rho^S_k(\tau) \rho^S_{-k}(\tau) 
\}}{{\mbox{Tr}} \bar H_{dz}^2},
\end{eqnarray}
where $\rho^S_k(\tau)$ is the contribution from the matrix $\rho^S$  to the $k$th order coherence. One has to take into account that each block $\rho^S(\tau)$ is degenerated with the multiplicity $n_N(S)$ \cite{LL,DFFZ}:
\begin{eqnarray}
n_N(S)=\frac{N!(2 S+1)}{(\frac{N}{2} + S +1)!(\frac{N}{2} -S)!},\;\;0\le S\le \frac{N}{2}.
\end{eqnarray}
As a result, the observable intensities $J_k(\tau)$ ($-N\le k\le N$) are following \cite{DFFZ,DFFZ2}:
\begin{eqnarray}
J_k(\tau) =\sum_Sn_N(S) J_{k,S}(\tau).
\end{eqnarray}
Remember that the dimensionality of each block $\bar H_{MQ}^S$  of the MQ NMR 
Hamiltonian  $\bar H_{MQ}$ is $(2S+1)$. Taking into account the block degeneration we 
obtain the correct value for the matrix dimensionality of both the Hamiltonina $\bar H_{MQ}$  and the density matrix $\rho$ \cite{DFFZ}:
\begin{eqnarray}
\sum_S n_N(S) (2S+1)= 2^N,
\end{eqnarray}
which is valid for the system of $N$ interacting spin-1/2 particles.

The numerical algorithms describing the MQ NMR dynamics in the systems of equivalent spins with the thermal equilibrium initial state  in the strong external magnetic field
have been developed in \cite{DFFZ,DFFZ2,DFZ}. With minor corrections, these algorithms may be used for the simulation of the dynamics of the MQ NMR coherences in the spin system with the dipolar ordered initial state. In particular, the integral of motion related with the MQ NMR Hamiltonian invariance with respect to the rotation over the angle $\pi$ around $z$-axis \cite{DFGM} is also present. Thus, for the odd $N$, it is enough to solve the problem for the MQ NMR Hamiltonians with two times lower matrix dimensionality and then double the resulting intensities  \cite{DFGM}.   We use the spin systems with odd $N$ in all numerical calculations in the next section.

\section{The numerical analysis of the MQ NMR profiles}
\label{Section:numerics}

\begin{figure*}
   \epsfig{file=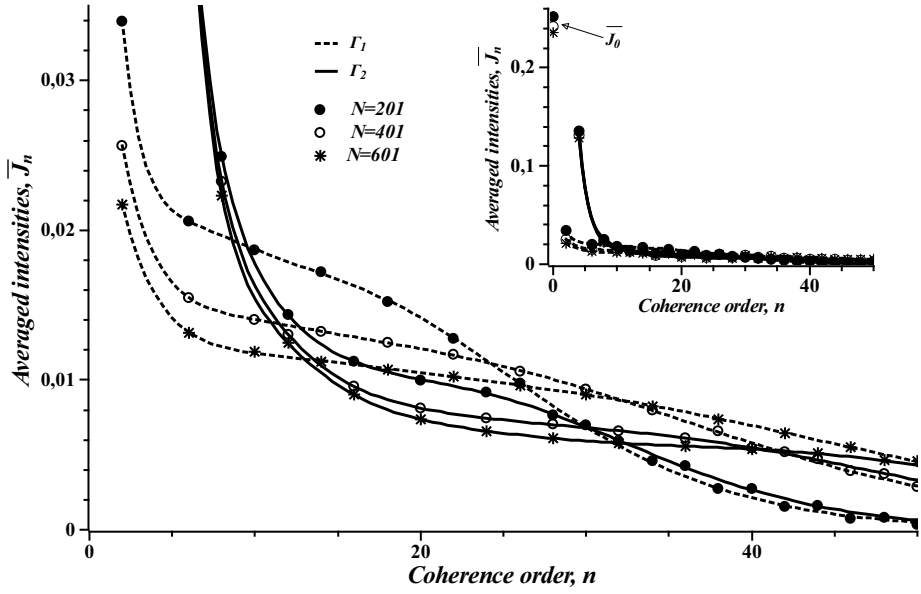
   , scale=0.5,angle=270
}
  \caption{Coherence intensities profiles for spin systems with $N=201,401,601$. The inset demonstrates that zero order coherence does not belong neither to $\Gamma_1$ nor to $\Gamma_2$. Only intensities of positive order coherences are presented. }
  \label{Fig:prof} 
\end{figure*}

Using the method developed in the previous section we investigate  the profiles of MQ NMR coherences. As far as all spins are ''nearest neighbors'' in the system of equivalent spins, $N$ spin cluster appears already after the time interval $\tau\sim 1/D$. However, some reorganization of this cluster is required for the MQ coherence formation \cite{DFFZ}.  The analysis of the MQ NMR coherence dynamics demonstrates that the quasistationary profile of MQ NMR coherences is created during  {$\bar \tau\sim 2$} ($\bar\tau=D\tau$ is the dimensionless time hereafter) and remains fast oscillating  for $\bar\tau>2$. Because of these oscillations it is convenient to use the averaged intensities $\bar J_k$ \cite{DFFZ} instead of intensities $J_k$ itselves. We estimate the dimensionless averaging time interval as $T\sim 2\pi /|\lambda^{min}_{3/2}|\approx 7.255 $, where $\lambda^{min}_{3/2}=\sqrt{3}/2$ is the minimal eigenvalue of the Hamiltonian \cite{DFFZ,DFFZ2}. For our convenience, we take $T=8$ so that 
\begin{eqnarray}\label{barI}
\bar J_k =\frac{1}{T}\int_{2}^{2+T} J_k(\bar\tau) d\bar\tau.
\end{eqnarray}
It is observed that $\bar J_k$ do not significantly vary with increase in $T$, so that the definition of the averaged intensities given by eq. (\ref{barI}) is valid. Although the dynamics of all coherences has been found in the numerical simulations, the intensities of the high order coherences are negligible, so that we represent  the intensities of the MQ NMR coherences  up to the 50th order in all figures below.   The  profiles of MQ NMR coherence intensities for systems of 201, 401 and 601 spins with dipolar ordered initial state  are shown in Fig.\ref{Fig:prof}.  These profiles are similar to those which have been found for the systems with the thermal equilibrium initial state in the strong external magnetic field \cite{DFFZ,DFFZ2}. Similar to refs.\cite{DFFZ,DFFZ2}, the averaged intensities of MQ NMR coherences are separated into two families:
\begin{eqnarray}\label{fam}
&&
\Gamma_1=\{\bar J_{4 k-2},\;\;k=0,\pm 1,\pm 2,\dots\}, \\\nonumber
&&
\Gamma_2=\{\bar J_{4 k},\;\;k=\pm 1,\pm 2,\dots\},
\end{eqnarray} 
with the zero order coherence  intensity $\bar J_0$ does not corresponds to any of these families.
Each family may be approximated by the smooth distribution function as follows:
\begin{eqnarray}\label{distr}
\bar J_{2k}\approx \left\{
\begin{array}{ll}
\displaystyle A_1\Big(1+\sum_{i=1}^4 (-1)^i a_{1i} (2 |k|)^i\Big) e^{-2\alpha_1  |k|}, &k=\pm 1,\pm 3,\dots\cr\displaystyle
A_2\Big(1+\sum_{i=1}^6(-1)^i a_{2i} (2 |k|)^i\Big) e^{-2\alpha_2  |k|}, &k=\pm 2,\pm 4,\dots\cr
\end{array}
\right.,
\end{eqnarray}
where parameters $A_i$, $a_{ij}$ and $\alpha_i$  for spin systems with $N=201$, 401 and 601 are collected in  Table 1.
The algorithm defining the approximation parameters is similar to that suggested in ref.\cite{DFFZ}.

\begin{table}[!htb]
\begin{tabular}{|c|c|c|c|c|c|c|}
\hline
N  &$A_1$                &$a_{11}$             & $a_{12}$            & $a_{13}$            & $a_{14}$            &$\alpha_1$\\\hline
201&$8.498\times 10^{-2}$&$3.133\times 10^{-1}$&$8.025\times 10^{-2}$&$6.109\times 10^{-3}$&$2.739\times 10^{-4}$&$2.437\times 10^{-1}$\\
401&$4.546\times 10^{-2}$&$1.694\times 10^{-1}$&$3.357\times  10^{-2}$&$1.649\times 10^{-3}$&$5.739\times 10^{-5}$&$1.638\times 10^{-1}$\\
601&$3.546\times 10^{-2}$&$1.456\times 10^{-1}$&$2.471\times 10^{-2}$&$1.028\times 10^{-3}$&$2.855\times  10^{-5}$&$1.342\times 10^{-1}$\\\hline
\end{tabular}
\begin{tabular}{|c|c|c|c|c|c|c|}
\hline
N  &$A_2$                &$a_{21}$             & $a_{22}$            & $a_{23}$            & $a_{24}$   & $a_{25}$  \\\hline
201&$2.217$\hspace*{1.3cm}&$4.029\times 10^{-1}$&$7.522\times 10^{-2}$&$7.638\times 10^{-3}$&$4.624\times 10^{-4}$&$1.562\times 10^{-5}$ \\
401&$1.411$\hspace*{1.3cm}&$3.587\times 10^{-1}$&$5.648\times  10^{-2}$&$4.616\times 10^{-3}$&$2.116\times 10^{-4}$&$5.136\times 10^{-6}$ \\
601&$1.210$\hspace*{1.3cm}&$3.441\times 10^{-1}$&$5.123\times 10^{-2}$&$3.924\times 10^{-3}$&$1.664\times  10^{-4}$&$3.686\times 10^{-6}$ \\\hline
\end{tabular}
\begin{tabular}{|c|c|c|}
\hline
N   & $a_{26}$           &$\alpha_2$\\\hline
201&$2.510\times 10^{-7}$&$3.039\times 10^{-1}$\\
401&$5.533\times 10^{-8}$ &$2.183\times 10^{-1}$\\
601&$3.512\times 10^{-8}$ &$1.940\times 10^{-1}$\\\hline
\end{tabular}
\label{Table:HPST2}
\caption{  The parameters $A_i$, $a_{ij}$ and $\alpha_i$   of the distribution function given by eq.(\ref{distr}) for  $N=201$, 401 and 601}
\end{table}

Similar to the profiles of the MQ NMR coherence intensities obtained for the systems of equivalent spin-1/2 particles with the thermal equilibrium initial state in the strong external magnetic field \cite{DFFZ},
profiles for the systems  with the dipolar ordered initial state seemed out to be exponential. This conclusion agrees with results obtained during an elaboration of numerous MQ NMR spectra \cite{LHG} and contradicts the phenomenological theory \cite{BMGP}  predicting the Gauss profiles of the MQ NMR coherence intensities.

Liouville equation (\ref{Liouville}) and formulas for the MQ NMR coherence intensities (\ref{Jk}) yield the conservation law  of the sum of the coherence intensities  \cite{LHGl}:
\begin{eqnarray}
\sum_k \bar J_k(\tau)=1.
\end{eqnarray}
This law together with the approximating formula (\ref{distr}) allows one to find a good approximation to the zero order coherence intensity.  We compare the calculated values of $\bar J_0$ (see eq.(\ref{Jk})) with the values 
$\bar J_0^{appr}=1-2 \sum_{k=1}^\infty J_{2k}$ found using the above conservation law and distribution function (\ref{distr}):
\begin{eqnarray}
\bar J_0=\left\{\begin{array}{ll}
2.519 \times 10^{-1},& N=201\cr
2.417 \times 10^{-1},& N=401\cr
2.361 \times 10^{-1},& N=601
\end{array}\right.,\;\;\;\bar J_0^{appr}=\left\{\begin{array}{ll}
2.501 \times 10^{-1},& N=201\cr
2.381 \times 10^{-1},& N=401\cr
2.387 \times 10^{-1},& N=601
\end{array}\right. .
\end{eqnarray}
Some discrepancy between $\bar J_0$ and $\bar J_0^{appr}$ appears because we take into account only coherences up to the 50th order in  constructing the distribution function (\ref{distr}), while contributions from the higher order coherences are missed.

\section{Comparison of the MQ NMR dynamics in the spin systems with 
two different initial states}
\label{Section:comp}

\begin{figure*}
   \epsfig{file=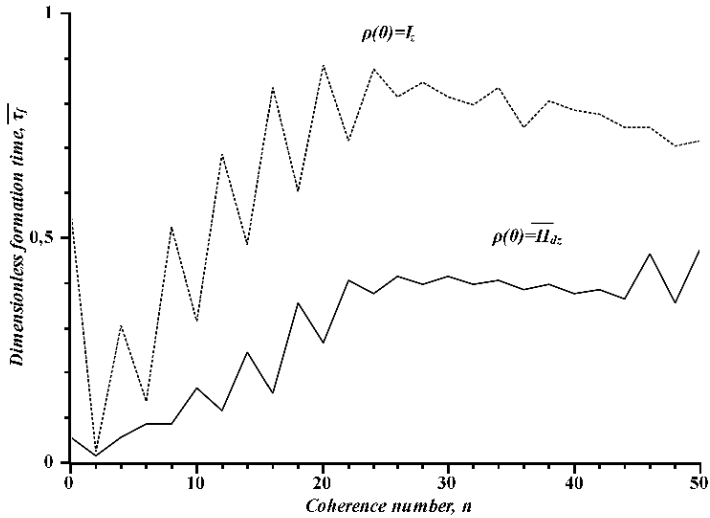
   , scale=0.5,angle=270
}
  \caption{Coherence formation time $\bar\tau_f$ versus the coherence number for the dynamics of  equivalent  spins with the dipolar ordered  and the thermal equilibrium in the strong external field  initial states.}
  \label{Fig:comp} 
\end{figure*}

The preparation of the system in the dipolar ordered initial state means that the two-spin correlations appear  already at the initial time, unlike the standard MQ NMR experiment, where   the thermal equilibrium initial state is defined by the one-spin Zeemann interaction with the external magnetic field \cite{DFKFG,DFKFG2}. This statement is justified by  Fig.\ref{Fig:comp}, where the formation times {$\bar\tau_f(n)$}  of different order coherences for both initial states are represented. Here the term  formation time  $\bar\tau_f(n)$ of  the $n$th coherence  means the time moment when the $n$th coherence intensity $J_n(\bar\tau)$ reaches the value  $\bar J_n$ for the first time. We see that MQ NMR coherences in the system with the dipolar ordered initial state (the lower solid line) appear much earlier. This result agrees with that obtained in \cite{DFKFG} for the MQ NMR in systems with small number of spin-1/2 particles.
\begin{figure*}
   \epsfig{file=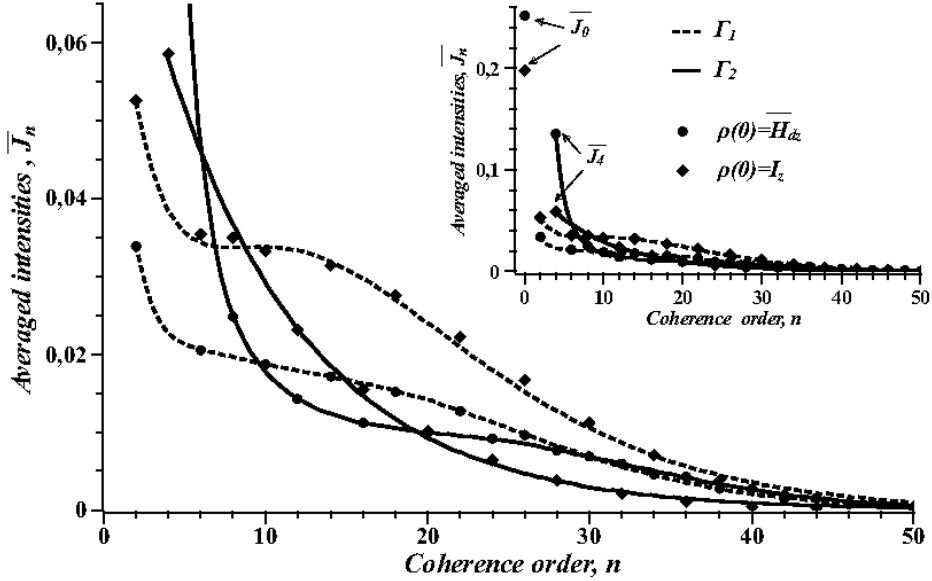
   , scale=0.6,angle=270
}
  \caption{The MQ NMR profiles for the systems of equivalent  spins with the  dipolar ordered and thermal equilibrium in the strong external magnetic field  initial states, $N=201$; the inset demonstrates that $\bar J_0$ and $\bar J_4$ are essentially bigger in the case of the dipolar ordered initial state.}
  \label{Fig:comp_prof} 
\end{figure*}

The profiles of the MQ NMR coherence intensities for  the system of $N=201$ spins with both the dipolar ordered  and the thermal equilibrium in the strong  magnetic field  initial states are compared in Fig.\ref{Fig:comp_prof}. This figure demonstrates that the discrepancy between two families of MQ NMR coherences $\Gamma_1$ and $\Gamma_2$ (see Eq.(\ref{fam})) is bigger for $n<10$ and smaller for $n>10$ in the case of the dipolar ordered initial state. The inset shows that $\bar J_0$ and $\bar J_4$ are 
essentially bigger in the case of the dipolar ordered initial state.

Thus, similar to the usual NMR  experiments in solids \cite{G}, the MQ NMR experiment in the systems of equivalent spins with the dipolar ordered initial state  may be usefull to supplement the MQ NMR experiment with the thermal equilibrium initial state  in the strong external magnetic field.

\section{Conclusions}
\label{Section:conclusion}
We have studied the MQ NMR  dynamics in the systems of equivalent spin-1/2 particles with the dipolar ordered initial state. For this purpose we modify the method developed in ref.\cite{DFFZ} for the system of equivalent spin-1/2 particles with the thermal equilibrium initial state in the strong external magnetic field. Similar to ref.\cite{DFFZ}, the high symmetry of such systems  allows one to investigate the dynamics in large spin systems containing hundreds of interacting spins. 
We obtain dependence of the MQ NMR coherence intensities  on their order (the profiles of the MQ NMR coherence intensities) in systems of 200-600 spins and demonstrate that these profiles may be well approximated by the exponential distribution functions. As far as the analogues result has been obtained in refs.\cite{DFFZ,DFFZ2} for the systems with the thermal equilibrium initial state in the strong external magnetic field, we may suppose that the exponential profiles of MQ NMR coherence intensities are  fundamental fact in MQ NMR dynamics.
The theoretical results obtained in refs.\cite{ZL,ZL2} confirm this conclusion. 

We demonstrate that the MQ NMR coherences appear faster in the spin systems with the dipolar ordered initial state.  The MQ NMR experiments with the dipolar ordered initial states expand possibilities of the MQ NMR spectroscopy in study of the structures of solids and the dynamical processes therein.

  All numerical simulations have been performed using the resources of the Joint Supercomputer Center (JSCC) of
the Russian Academy of Sciences.  The work was
supported by the Program of the Presidium of Russian Academy of Sciences No.21 " Foundations of fundamental
investigations of nanotechnologies and nanomaterials".


\begin{thebibliography}{99}

\bibitem{BMGP}
J.Baum, M.Munovitz, A.N.Garroway and A.Pines, J. Chem. Phys., {\bf 83}, 2015 (1985).

\bibitem{BGPGR}
J.Baum, K.K.Gleason, A.Pines, A.N.Garroway and J.A.Reimer, Phys.Rev.Lett. {\bf 56}, 1377 (1986).

\bibitem{H}
C.E.Hughes, Prog. Nucl. Magn. Reson. Spectrosc. {\bf 45}, 301 (2004)

\bibitem{WWP}
W.S.Warren, D.P.Weitekamp and A.Pines, J. Chem. Phys. {\bf 73}, 2084 (1980)

\bibitem{FG}
G.B.Furman and S.D.Goren, J.Phys.:Condens.Matter {\bf 17}, 4501 (2005)

\bibitem{G}
M. Goldman, Spin Temperature and Nuclear Magnetic Resonance in Solids (Clarendon, Oxford, 1970).

\bibitem{SH}
C.P.Slichter and W.C.Holton, Phys.Rev.  {\bf 122}, 1701 (1961)

\bibitem{JB}
J.Jeener and P.Broekaert, Phys. Rev. {\bf 157}, 232 (1967)

\bibitem{DFKFG}
S.I.Doronin, E.B.Fel'dman, E.I.Kuznetsova, G.B.Furman and S.D.Goren,
 Phys.Rev.B {\bf 76}, 144405 (2007)


\bibitem{DFKFG2}
S.I.Doronin, E.B.Fel'dman, E.I.Kuznetsova, G.B.Furman and S.D.Goren,
 JETP Letters {\bf 86}, 24  (2007)

\bibitem{DRKN}
V.V.Dobrovitski, H.A.De Raedt, M.I.Katsnelson and B.N.Harmon, Phys.Rev.Lett. {\bf 90}, 210401 (2003)

\bibitem{ZCAPCDRV}
W.X.Zhang, P.Cappellaro, N.Amtler, B.Pepper, D.G.Cory, V.V.Dobrovitski, C.Ramanathan and L.Viola, Phys. Rev. A {\bf 80}, 052323 (2009)

\bibitem{BKHWW}
J.Baugh, A.Kleinhammes, D.Han, Q.Wang and Y.Wu, Science {\bf 294},  1505 (2001)

\bibitem{FR}
E.B.Fel'dman and M.G.Rudavets, J.Exp.Theor.Phys. {\bf 98}, 207 (2004)

\bibitem{DFFZ}
S.I.Doronin, A.V.Fedorova, E.B.Fel'dman and A.I.Zenchuk, J.Chem.Phys. {\bf 131} 104109 (2009)

\bibitem{DFFZ2}
S.I.Doronin, A.V.Fedorova, E.B.Fel'dman and A.I.Zenchuk, Phys.Chem.Chem.Phys. {\bf 12}, 13273 (2010)

\bibitem{DFZ}
S.I.Doronin, E.B.Fel'dman and A.I.Zenchuk,  J.Chem.Phys. {\bf 134}, 034102 (2011)

\bibitem{D}
F.S.Dzheparov, Preprint ITEP No. 7-10 (2010)

\bibitem{FL}
E.B.Fel'dman and S.Lacelle, J.Chem.Phys. {\bf 107}, 7067 (1997)

\bibitem{LL}
L. D. Landau and E. M. Lifshitz, Course of Theoretical Physics, Vol. 3: Quantum Mechanics: Non-Relativistic Theory
(Nauka, Moscow, 1974; Pergamon, New York, 1977).

\bibitem{DFGM}
S.I.Doronin, E.B.Fel'dman, I.Ya.Guinzbourg and I.I.Maximov, Chem. Phys. Lett. {\bf 341}, 144 (2001).


\bibitem{LHG}
S.Lacelle, S.-J.Hwang and B.C.Gerstein, J.Chem.Phys. {\bf 99}, 8407 (1993)

\bibitem{LHGl}
D.A.Lathrop, E.S.Handy and K,K,Gleason, J.Magn.Reson. Ser.A {\bf 111}, 161 (1994)

\bibitem{ZL}
V.E.Zobov and A.A.Lundin, J.Exp.Theor.Phys. {\bf 103}, 904 (2006)

\bibitem{ZL2}
V.E.Zobov and A.A.Lundin, J.Exp.Theor.Phys. {\bf 139}, 519 (2011)


\end{thebibliography}
\end{document}